\documentclass[aps,pre,twocolumn,showpacs,amsmath,amssymb,floatfix]{revtex4}

\usepackage{graphicx}
\usepackage{epsfig}

\begin{document}

\title{Global persistence exponent of the double-exchange model}
\author{H. A. Fernandes\footnote{Email address: henrique@pg.ffclrp.usp.br} and
J. R. Drugowich de Fel\'{\i}cio\footnote{Email address: drugo@usp.br}}
\affiliation{Departamento de
F\'{\i}sica e Matem\'{a}tica, Faculdade de Filosofia, Ci\^{e}ncias e
Letras de Ribeir\~{a}o Preto, Universidade de S\~{a}o Paulo, Avenida
Bandeirantes, 3900 - CEP 14040-901, Ribeir\~{a}o Preto, S\~{a}o Paulo,
Brazil}

\begin{abstract}

  We obtained the global persistence exponent $\theta_g$ for a
  continuous spin model  on the simple cubic lattice with
  double-exchange interaction by using two different methods. First,
  we estimated the exponent $\theta_g$ by following the time evolution
  of probability $P(t)$ that the order parameter of the model does not
  change its sign up to time $t$ $[P(t)\thicksim t^{-\theta_g}]$.
  Afterwards, that exponent was estimated through the scaling collapse
  of the universal function $L^{\theta_g z} P(t)$ for different
  lattice sizes. Our results for both approaches are in very good
  agreement each other.

\end{abstract}

\pacs{64.60.Ht, 75.47.Gk, 02.70.Uu}
\date{\today}

\maketitle

During the last few years, a great number of experimental and
theoretical works have been published about manganites with perovskite
structure, $A_{1-x}B_x\mbox{MnO}_3$, where $A$ is a rare earth such as
La, Nd, and Pr, $B$ is a divalent element such as Sr, Ca, Ba, and Pb,
and $x$ is the concentration of $B$. Such attention is due to the
rediscovery of colossal magnetoresistence
(CMR) \cite{Helmolt1993,Ramirez1997,Coey1999,Salomon2001}, an extremely
large change in the resistivity when a magnetic field is applied near
of Curie temperature. Besides, these manganese oxides possesse
metal-insulator (MI) transitions, as well as a rich variety of
physical properties and possible technological applications, for
instance, magnetic sensors and memory technology.

In these compounds, the MI transition associated with the
ferromagnetic spin alignment has been widely explained by the
double-exchange (DE) mechanism
\cite{Zener1951,Anderson1955,Gennes1960,Varma1996,
Furukawa1994,Goodenough1955}, in which the transfer of an itinerant
$e_g$ electron between the neighboring Mn ions through the
$\mbox{O}^{2-}$ ion results in a ferromagnetic interaction. Anderson e
Hasegawa \cite{Anderson1955} showed that the transference element is
proportional to $\cos(\phi/2)$, where $\phi$ is the angle between
the ionic neighbouring spins. This result was recently confirmed for
layered manganites \cite{Qingan2000}. Although this theory has also
been succeeded in explaining qualitatively CMR, some authors have
argued that it cannot alone provide a complete description of this
phenomenon.  They suggest that, in addition to the double-exchange, a
complete understanding of these materials should include strong
electron correlations \cite{Asamitsu1996}, a strong electron-phonon
interaction \cite{Millis1995}, or coexisting phases
\cite{Lynn1996}. One might therefore think that double-exchange alone
cannot explain CMR in manganites \cite{Millis1995}, but this remains
an open question. What we know is that in the study of the manganites
the double-exchange theory plays an important role, both in the study
of CMR and in explaining the presence of a ferromagnetic state (for $x
\thickapprox 0.3$) in doped manganites, furnishing the basis for
describing manganites with colossal magnetoresistence.

The Hamiltonian of a classical spin model with double-exchange
interactions is given by \cite{Caparica2000}
\begin{equation}
\mathcal{H}=-J\sum_{<i,j>}\sqrt{1+\textbf{S}_i\cdot\textbf{S}_j}
\label{hde},
\end{equation}
where $\langle i,j\rangle$ indicates that the sum runs over all nearest-neighbor
pairs of lattice site, $J$ is the ferromagnetic coupling constant and
the spin $\textbf{S}_i=(S_i^x,S_j^y,S_k^z)$ is a three-dimensional
vector of unit length.

The critical properties of the DE models have been intensively studied
by using Monte Carlo (MC) simulations. In the equilibrium
\cite{Caparica2000,Motome2000,Motome2001,Motome2003}, the estimates
for the static critical exponents indicate that this model belongs to
the universality class of the classical Heisenberg model
\cite{Chen1993}. Several experimental works about the critical
properties of the doped perovskite manganites also support this
assertion \cite{Ghosh1998,Sahana2003,Li2004}. Very recently the
dynamic critical behavior of the DE model was studied by using the
short-time Monte Carlo simulations and estimates for the static
critical exponents $\nu$ and $\beta$ and the dynamic critical
exponents $z$ and $\theta$ were derived \cite{Fernandes2005}. That
approach is based on the results of Janssen \textit{et al.}
\cite{Janssen1989} that showed that
universality and scaling behavior are already present in systems since
their early stages of the time evolution after quenching them from
high temperatures to the critical one. By using renormalization group
techniques they obtained for the \textit{k}-th
moment of the magnetization, extended to systems of finite size
\cite{Li1995}, the following scaling relation
\begin{equation}
M^{(k)}(t,\varepsilon ,L,m_{0}) =b^{-k\beta /\nu}M^{(k)}(b^{-z}t,b^{1/\nu}\varepsilon
,b^{-1}L,b^{x_{0}}m_{0}) \label{eq1},
\end{equation}

where $t$ is time, $b$ is an arbitrary spatial rescaling factor,
$\varepsilon=\left(T-T_{c}\right)/T_{c}$ is the reduced temperature
and $L$ is the linear size of the lattice. The exponents $\beta$ and
$\nu$ are the equilibrium critical exponents associated with the
order parameter and the correlation length, and $z$ is the dynamical
exponent ($\tau \thicksim  \xi^z $ where $\tau$ is the time
correlation). For a large lattice size $L$ and small initial
magnetization $m_0$ at the critical temperature $(\varepsilon=0)$,
the magnetization is governed by a new dynamic exponent $\theta$,
\begin{equation}
M(t) \thicksim m_0t^\theta \label{eq2}
\end{equation}
if we choose the scaling factor $b=t^{1/z}$ in Eq. (\ref{eq1}). This
new critical index, independent of the previously known exponents,
characterizes the so-called ``critical initial slip'', the anomalous
increase of the magnetization when the system is quenched to the
critical temperature $T_c$. In the sequence, another
new independent dynamic critical exponent was found
by Majumdar \textit{et al.} \cite{Majumdar1996} to describe the
behavior of the global persistence probability $P(t)$ that the order
parameter has not changed its sign up to time $t$. At criticality,
$P(t)$ is expected to decay algebraically as
\begin{equation}
P(t) \thicksim t^{-\theta_g},\label{Eq:thetag}
\end{equation}
where $\theta_g$ is the global persistence exponent. Since then, the
study of the persistence behavior have attracted an enormous interest,
playing an important role in the study of systems far from equilibrium
\cite{Majumdar1996,Majumdar2003,Schulke1997,Oerding1997,Dasilva2003,Dasilva2005,
Ren2003,Albano2001,Saharay2003,Hinrichsen1998,Sen2004,Zheng2002,Cueille1999}.

Since the time evolution of the order parameter is in general a
non-Markovian process, the new critical exponent $\theta_g$ is
independent of the usual exponents. However, as argued by
Majumdar \textit{et al.} \cite{Majumdar1996}, if the global order
parameter is described by a Markovian process, $\theta_g$ is not a new
independent exponent, being related to other critical exponents by the equation

\begin{equation}
\theta_g=\alpha_g=-\theta+\frac{d}{z^2}-\frac{\beta}{\nu z}. \label{Eq:scarel}
\end{equation}

The global persistence probability $P(t)$ can be defined as
\begin{equation}
P(t)=1-\sum_{t'=1}^t \rho(t'),
\end{equation}
where $\rho(t')$ is the fraction of the samples that have changed
their state for the first time at the instant $t'$.

In this paper, we performed short-time Monte Carlo simulations to
explore the scaling behavior of the global persistence probability
$P(t)$, for a classical ferromagnet with double-exchange interaction.
The dynamical exponent $\theta_g$ that governs the behavior of $P(t)$
at criticality is obtained by using two different approaches: the
straight application of the power law behavior [see
Eq. (\ref{Eq:thetag})] and by means of time series data collapse. To
our knowledge, this is the first time that this exponent is calculated
for a three dimensional model and continuous spin variables.

In our simulations, we considered $L\times L \times L$ ($L=20$, 25,
30, 35, 40, 50 and 60) simple cubic lattices with periodic boundary
conditions. Simulations were done at critical temperature
\cite{Caparica2000} $T_c=0.74515$, in units of $J/k_B$, where $k_B$
is the Boltzmann's constant.  The update we used is local and
follows the Metropolis algorithm, i.e., at each site of the lattice
(during the simulation) a trial orientation of the spin is randomly
chosen and accepted or rejected according to the probability
$\mbox{e}^{-\beta(E'-E)}$ where $E'$ $(E)$ is the new (old) energy
of the spin system, $\beta=J/k_BT_c$ and  $k_B$ is the Boltzmann
constant. The estimates were obtained from five independent bins
with 5000 samples each for $t$ up to 500 Monte Carlo sweeps.

In the first method, we used the scaling relation given by Eq.
(\ref{Eq:thetag}) in order to obtain $\theta_g$ as a function of $m_0$
for several values of the initial magnetization. Here, it is necessary
working with a precise and small value of the  magnetization,
$m_0 \thicksim 0$. Next, the estimate for $\theta_g$ is obtained
extrapolating that series to the limit in which $m_0\rightarrow 0$.

\begin{figure}[ht]
\centering \epsfig{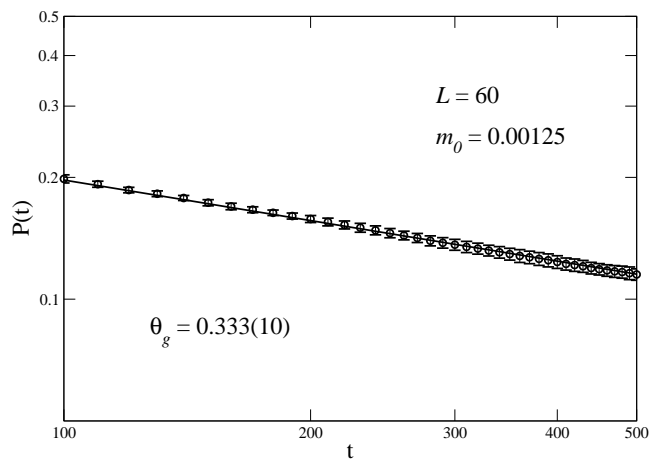}
\caption{Time evolution of the global persistence probability $P(t)$
for a lattice size $L=60$ and $m_0=0.00125$. The error bars were
calculated over 5 sets of 5000 samples.}
\label{Fig:limit60}
\end{figure}

\begin{figure}[ht]
\centering \epsfig{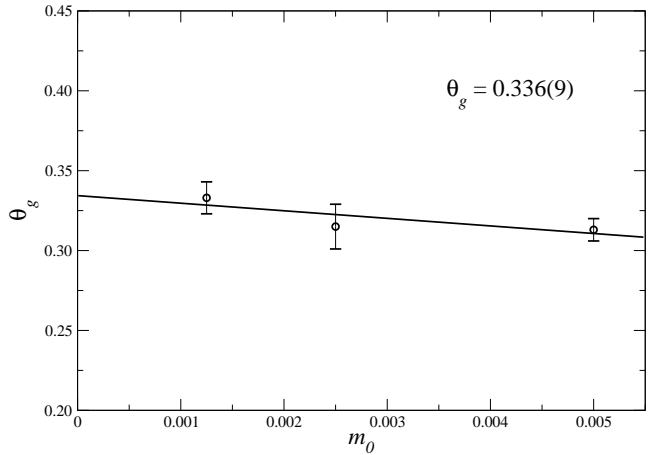}
\caption{Dynamical exponent $\theta_g$ as a function of the initial
magnetization $m_0$ for cubic lattices with $L=60$. Each point represents
an average over 5 sets of 5000 samples.}
\label{Fig:extrapolation}
\end{figure}

In Fig. \ref{Fig:limit60} we show the decay of the global
persistence probability for $L=60$ and $m_0=0.00125$ in double-log
scales.
Fig. \ref{Fig:extrapolation} exhibits the behavior of the exponent
$\theta_g$ for $m_0=0.005$, 0.0025 and 0.00125, as well as a linear
fit that leads to the value $\theta_g=0.336(9)$.

\begin{table*}[!htb]\centering
\caption{The global persistence exponent $\theta_g$ from the power
law behavior for different initial magnetizations $m_0$ and even lattice sizes.}
\label{Table:limit}
\begin{tabular}{c c c c c}
\hline\hline ~~~ $L$ &~~~~~ $m_0=0.005$ &~~~~~ $m_0=0.0025$ &~~~~~ $m_0=0.00125$ &~~~~~ Extrapolated value ~~~ \\
\hline ~~~20 &~~~~~ 0.315(14) &~~~~~ 0.326(8)  &~~~~~ 0.337(12) &~~~~~ 0.342(4)~~~ \\
       ~~~30 &~~~~~ 0.315(5)  &~~~~~ 0.322(12) &~~~~~ 0.335(9)  &~~~~~ 0.339(5)~~~ \\
       ~~~40 &~~~~~ 0.315(9)  &~~~~~ 0.321(31) &~~~~~ 0.336(28) &~~~~~ 0.339(7)~~~ \\
       ~~~50 &~~~~~ 0.319(10) &~~~~~ 0.325(11) &~~~~~ 0.336(14) &~~~~~ 0.338(6)~~~ \\
       ~~~60 &~~~~~ 0.313(7)  &~~~~~ 0.315(14) &~~~~~ 0.333(10) &~~~~~ 0.336(9)~~~ \\
\hline\hline
\end{tabular}
\end{table*}

In Table \ref{Table:limit} we present the estimates for $\theta_g$ in
function of different initial magnetizations $m_0$ for other lattice
sizes.

Our results obtained through the linear extrapolations for
$m_0\rightarrow 0$ are presented in the last column. The estimate of
the exponent $\theta_g$ for odd lattice sizes ($L=25$ and $L=35$) were
obtained for different values of $m_0$. For $L=25$ we used
$m_0=0.0064$, 0.0032 and 0.0016 and the extrapolated value was
$\theta_g=0.337(5)$, whereas for $L=35$ we used $m_0 = 240/(35^3)$,
$120/(35^3)$, and $60/(35^3)$ and the extrapolated value
$\theta_g=0.338(5)$.

In the second method, we used the fact that the dependence of $P(t)$
on the initial magnetization can be cast in the following
finite-size scaling relation \cite{Majumdar1996}
\begin{equation}
P(t)=t^{-\theta_g}f(t/L^z)=L^{-\theta_g z}\tilde{f}(t/L^z), \label{Eq:p2}
\end{equation}
where $z$ is the dynamical exponent. Thus, the quantity $L^{\theta_g
  z} P(t) $ is an universal function of the scaled time ($t/L^z$) and
the wanted value of $\theta_g$ is that which fulfils that condition
for different lattices. The best estimate for $\theta_g$ is found
through the $\chi^2$ test \cite{Press1986}.

Unlike the first method used in this paper, in the collapse method the
exponent $\theta_g$ is obtained without the need of careful
preparation of the initial magnetization $m_0$ nor the limiting
procedure. The only requirement is that $\langle m_0 \rangle \thicksim
0$ where $\langle \cdot \rangle$ is an average done over the
samples at $ t = 0$. On the other hand, the collapse method demands the dynamical
exponent $z$ to be known beforehand. In this paper, we used the
estimate obtained very recently for this exponent, $z=1.975(10)$
\cite{Fernandes2005}.

\begin{figure}[ht]
\centering \epsfig{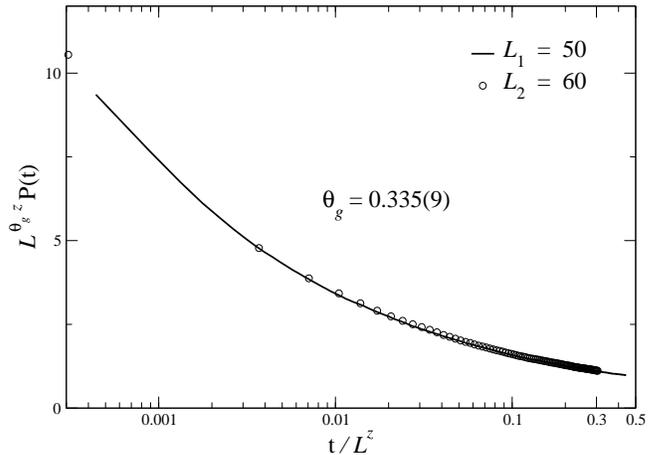}
\caption{Scaling Collapse of the scaled persistence probability versus
  scaled time for $L_1 = 50$ and $L_2 = 60$. The curves were obtained from
  five independent bins of 5000 samples.}
\label{Fig:collapse}
\end{figure}

In Fig. \ref{Fig:collapse} we show the collapse of the curves
obtained for $L=50$ and $L=60$.
The open circles show the collapse of the larger lattice rescaled in
time. Our best estimate for $\theta_g$ for $L_2=60$ and $L_1=50$ is
\begin{equation}
\theta_g=0.335(9) \label{Eq:collapse}
\end{equation}
The estimates for other lattice sizes are shown in Table
\ref{Table:collapse}.
\begin{table}[!ht]\centering
\caption{The global persistence exponent $\theta_g$ for the best
collapse for the double-exchange model.} \label{Table:collapse}
\begin{tabular}{c c}
\hline\hline $L_2 \longmapsto L_1$ &~~~~ $\theta_g$ \\
\hline $60 \longmapsto 50$ &~~~~ 0.335(9)  \\
       $60 \longmapsto 40$ &~~~~ 0.333(13) \\
       $60 \longmapsto 30$ &~~~~ 0.329(14) \\
       $50 \longmapsto 35$ &~~~~ 0.337(11) \\
       $40 \longmapsto 25$ &~~~~ 0.330(12) \\
       $40 \longmapsto 20$ &~~~~ 0.332(12) \\
\hline\hline
\end{tabular}
\end{table}
These results are in very good agreement with the
estimates obtained directly from the power law predicted in
Eq. (\ref{Eq:thetag}).

Using, for instance, the result of Eq. (\ref{Eq:collapse}) and the
estimates of the exponents $\theta$, $z$, $\beta$, and $\nu$
obtained in Ref. \cite{Fernandes2005}, both for the largest lattice
size ($L=60$), we verify through Eq. (\ref{Eq:scarel}), the
non-Markovian aspect of the phenomenon we are deal with, since
\begin{equation}
\theta_g=0.335(9) \hspace{0.2cm} \mbox{and} \hspace{0.2cm}
\alpha_g=-\theta  + \frac{d}{z^2} - \frac{\beta}{\nu z}=0.026(17).
\label{Eq:nonMarkovian}
\end{equation}

Thus, the global persistence exponent is also in this case
independent of other critical exponents but the difference between
our estimate for $\theta_g$ and the value obtained from Eq.
(\ref{Eq:scarel}) is greater than that observed when discrete spin
models were analyzed (see Table \ref{Table:results}).
\begin{table}[ht]\centering
  \caption{The exponents $\theta_g$ and $\alpha_g$ for several models.} \label{Table:results}
\begin{tabular}{c c c}
\hline\hline Models &~~~~ $\theta_g$ &~~~~ $\alpha_g$ \\
\hline Ising model \cite{Schulke1997} &~~~~ 0.236(3) &~~~~ 0.212(2) \\
       Three-state Potts model \cite{Schulke1997} &~~~~ 0.350(8) &~~~~ 0.324(3) \\
       Blume-Capel model \cite{Dasilva2003} &~~~~ 1.080(4) &~~~~ 0.904(21) \\
       DE model (see Eq.(\ref{Eq:nonMarkovian})) &~~~~ 0.335(9) &~~~~ 0.026(17) \\
\hline\hline
\end{tabular}
\end{table}

In summary, we have performed short-time Monte Carlo simulations in
order to investigate the scaling behavior of the persistence
probability $P(t)$ for a three dimensional system with double-exchange
interaction. The dynamic critical exponent $\theta_g$ that governs the
behavior of $P(t)$ at criticality was estimated by using two different
approaches: the straight application of the power law behavior $P(t)
\thicksim t^{-\theta_g}$ and the collapse method for the universal
function $L^{\theta_g z} P(t)$. The results are consistent with the
expected non-Markovian character of the process.

This work was
supported by the Brazilian agencies CAPES and CNPq.

\end{document}